\documentclass{elsart}
\usepackage{graphicx}
\usepackage{fleqn}
\begin{document}
\begin{frontmatter}
\title{Hysteresis phenomena at ultrathin lubricant film 
melting in the case of first-order phase transition}
\author{A.V.~Khomenko\thanksref{AH}},
\author{I.A.~Lyashenko}
\address{Sumy State University, Rimskii-Korsakov St. 2, 40007 Sumy, Ukraine}
\thanks[AH]{Corresponding author. \\ E-mail addresses: khom@phe.sumdu.edu.ua (A.V. Khomenko),
nabla@cable-tv.sumy.ua (I.A.~Lyashenko)}
\date{\today}
\begin{abstract}
Within the framework of Lorentz model for description of viscoelastic medium 
the influence of deformational defect of the shear modulus is studied on 
melting of ultrathin lubricant film confined between the atomically flat solid 
surfaces. 
The possibility of jump-like and continuous melting is shown. 
Three modes of lubricant behavior are found, which correspond to the zero 
shear stress, the Hooke section of loading diagram, and the domain of plastic 
flow. Transition between these modes can take place according to mechanisms of 
first-order and second-order phase transitions. 
Hysteresis of dependences of stationary stresses on strain and 
friction surfaces temperature is described. Phase kinetics of the system is 
investigated. 
It is shown that ratio of the relaxation times for the studied quantities 
influences qualitatively on the character of the stationary mode setting. 
\end{abstract}
\end{frontmatter}

\section{Introduction}\label{sec:level1}

Last years the problems of sliding friction of smooth solid surfaces with 
thin lubricant film between them attract considerable attention \cite{1lit}. 
Experimental study of atomically flat 
surfaces of mica, separated by ultrathin lubricant film, has 
shown that it manifests the properties of solid at certain conditions 
\cite{Yosh}. Particularly, the interrupted motion 
(stick-slip) was observed inherent to dry friction \cite{2lit,Aranson}. 
Such boundary mode is realized, if film of lubricating material has 
less than four molecular layers. This is explained as solidification 
conditioned by the compression of walls. Subsequent jump-like melting 
takes place, when shear stress exceeds a critical value due to 
"shear melting effect". 
These films are characterized by a yield stress, which is a 
characteristic of failure in solids. 

In the previous work \cite{liq} the approach is developed according to 
which the transition of ultrathin lubricant film from solid-like to 
liquid-like state takes place as a result of 
thermodynamic and shear melting. The combined analytical description of 
these processes is carried out due to self-organization of the fields 
of shear stresses and strain, and also temperature of lubricant film 
taking into account additive noises of the indicated quantities 
\cite{pla,stoh} and correlated fluctuations of temperature 
\cite{vest,vest05}. The influence of additive noises on phase kinetics of the 
system is analyzed \cite{vestKIN}. 

However, there a question was not considered about the reasons for the 
jump-like melting 
and hysteresis, which were observed in the experiments \cite{exp1}. 
The proposed work is devoted to finding of realization conditions of these 
features taking into account the deformational defect of the shear modulus. 
Our approach is based on the Lorentz model for approximation of viscoelastic 
medium 
\cite{liq} at the conditions when the first-order phase transition 
is realized. Three stationary mode are found --- 
two solid-like, corresponding to dry friction, and one liquid-like, 
meeting the sliding. 
The phase portraits of the system are built. The various interrupted modes 
of friction depending on initial conditions until setting of equilibrium 
can be realized. 
It is also shown that in the case, when relaxation time of 
strain much longer than corresponding times for stresses and temperature, 
the hysteresis is manifested in phase portraits. 

\section{Basic equations}\label{sec:level2}

In the previous work \cite{liq} on the basis of rheological description of 
viscoelastic medium, possessing heat conductivity, the system of kinetic 
equations is obtained, which determine the mutually coordinated behavior of 
shear stresses $\sigma$ and strain $\varepsilon$, and also 
temperature $T$ in ultrathin lubricant film during the process of friction 
between atomically flat mica surfaces. 
Let us write down these equations using the measure units: 
\begin{equation}
\label{eq1d}
\sigma_s = \left( {\frac{\rho c_v \eta _0 T_c }{\tau _T }} \right)^{1 / 
2}, \quad \varepsilon _s = \frac{\sigma_s}{G_0 } \equiv \left( {\frac{\tau 
_\varepsilon }{\tau _T }} \right)^{1 / 2}\left( {\frac{\rho c_v T_c \tau 
_\varepsilon }{\eta _0 }} \right)^{1 / 2}, \quad T_c
\end{equation}
for variables $\sigma,\varepsilon,T$, respectively, where $\rho $ is the 
mass density, $c_v $ is the specific heat capacity, $T_c$ is the critical 
temperature, $\eta _0 \equiv \eta \left( {T = 2T_c } \right)$ is the 
characteristic value of shear viscosity $\eta $, 
$\tau _T \equiv \rho l^2c_v / \kappa$ is the time of heat conductivity, 
$l$ is the scale of heat conductivity, 
$\kappa$ is the coefficient of heat conductivity, 
$ \tau_\varepsilon$ is the relaxation time of strain, 
$G_0 \equiv \eta _0 / \tau _\varepsilon$: 
\begin{eqnarray}
&&\tau _{\sigma}\dot{\sigma}=-\sigma + g(\sigma)\varepsilon, \label{eq1} \\
&&\tau_{\varepsilon}\dot{\varepsilon}=-\varepsilon + (T-1)\sigma,
\label{eq2} \\
&&\tau_{T}\dot{T}=(T_e - T) - \sigma \varepsilon + \sigma^2. \label{eq3}
\end{eqnarray} 
Here the relaxation time of stress $\tau_\sigma$, the 
temperature $T_e$ of atomically flat mica surfaces of friction, and the 
function $g(\sigma) \equiv G(\sigma) /G_0$ are introduced, where $G(\sigma)$ 
is the shear modulus of lubricant depending on the stress value: 
\begin{equation}
G(\sigma) = \Theta + \frac{G-\Theta}{1+(\sigma/\sigma_p)^\beta}, \quad 
\beta = {\rm const} > 0. 
\label{NN}
\end{equation}
At $g(\sigma) = G/G_0 \equiv {\rm const}$ equation (\ref{eq1}) is reduced to 
the Maxwell-type equation for description of viscoelastic medium by 
substituting 
$\varepsilon / \tau_{\sigma}$ for $\partial \varepsilon / \partial t$. 
The Maxwell equation supposes the use of the idealized Genky model. 
For the dependence of stress on strain $\sigma (\varepsilon )$ 
this model is represented by the Hooke law $\sigma=G\varepsilon $ at 
$\varepsilon < \varepsilon _m $ and constant $\sigma _m = G\varepsilon _m $ 
at $\varepsilon \ge \varepsilon _m $ ($\sigma _m $, $\varepsilon _m $ are the 
maximal values of shear stress and strain for the Hooke section, 
$\varepsilon > \varepsilon_m$ results in the viscous flow with deformation 
rate $\dot {\varepsilon } = (\sigma - \sigma _m )/\eta$). Actually, the 
curve of dependence $\sigma (\varepsilon )$ has two regions: 
first one, Hookean, has the large slope determined by the shear 
modulus $G$, and it is 
followed by the more gently sloping section of plastic deformation whose tilt 
is defined by the hardening factor $\Theta < G$. Obviously, 
the above picture means that the shear modulus depends on the value of 
stresses. 
For the account of this circumstance we use the simplest approximation 
(\ref{NN}), which describes the above represented transition of elastic 
deformation mode to the plastic one. It takes place at the characteristic 
values 
of shear stress $\sigma_p$ and strain $\varepsilon_p$. 
It is worth noting that at description of structural phase transitions of 
liquid-like lubricant the third-order invariants are present, which 
breaks the parity of the dependence of the synergetic potential $V$ on the 
stress 
$\sigma$. Therefore in approximation (\ref{NN}) the linear term was used 
$\sigma / \sigma_p $ ($\beta=1$), instead of square one  
$(\sigma / \sigma_p )^2$ ($\beta=2$) \cite{zhetph}, and dependence 
$V(\sigma)$ was not already even \cite{liq}. 

Expression (\ref{eq2}) has the form of the corresponding Kelvin-Voight 
equation 
\cite{liq,voigt}, which takes into account the dependence of the shear 
viscosity on the dimensionless temperature 
\begin{equation}
\eta = \frac{\eta_0}{T-1}.
\label{visc}
\end{equation}
Note that jointly equations (\ref{eq1}) and (\ref{eq2}) represent the new 
rheological model, since they are reduced to the second-order differential 
equation with respect to the stress $\sigma$ or the strain $\varepsilon$. 
Equation (\ref{eq3}) is the expression for heat conductivity, which 
describes the transmission of heat from the friction surfaces to lubricant 
film, the effect of dissipative heating of viscous liquid flowing under the 
action of stress, and the reversible mechanic-and-caloric effect in linear 
approximation. 
Equations (\ref{eq1}) -- (\ref{eq3}) formally coincide with the synergetic 
Lorentz system \cite{zhetph,Haken}, in which the shear stress acts as 
the order parameter, the conjugate field is reduced to the shear strain, and 
the temperature represents the control parameter. As is known this system is 
used for description of both phase thermodynamic and kinetic 
transformations. It is necessary also to note that rheological properties of 
lubricant film are investigated experimentally that allows to build a phase 
diagram \cite{Yosh}. 

Dependence (\ref{NN}) describes hysteresis at 
melting of thin layer of lubricant only in coordinates $T_e - \sigma$ during 
realization of first-order phase transition \cite{liq}. 
Thus, deformation curve $\sigma(\varepsilon)$ is monotonous and 
in the case of second-order phase transition it allows us to represent only 
continuous transition. However, experimental data testify that 
melting of molecularly thin lubricant film has jump-like character 
\cite{Yosh}, though may take place according to the mechanism of second-order 
phase transition. 
As is shown below the description of the indicated feature is achieved 
replacing dependence $g(\sigma)$ in (\ref{eq1}) by 
$g(\varepsilon) \equiv G(\varepsilon) /G_0$, where 
\begin{equation} 
G(\varepsilon) = \Theta + 
\frac{G-\Theta}{1+(\varepsilon/\varepsilon_p)^\beta}. 
\label{G}
\end{equation}
At this time the value of parameter $\beta > 0$, determining the potential 
character, plays a key role also. 

In the work \cite{liq} the melting of ultrathin lubricant film during the 
process of friction between the atomically flat surfaces of mica 
represented as a result of 
spontaneous appearance of shear stresses resulting in plastic flow. 
This is caused by heating of friction surfaces above the critical 
value $T_{c0} = 1+ G_0/G$. 
The initial reason for self-organization process is the positive feedback of 
$T$ and $\sigma$ on $\varepsilon$ [see (\ref{eq2})], conditioned by 
temperature dependence of shear viscosity leading to its 
divergence. On the other hand, the negative feedback of 
$\sigma$ and $\varepsilon$ on $T$ in (\ref{eq3}) has an important 
role since it provides the system stability. 

In accordance with such approach the lubricant represents a strongly viscous 
liquid, 
which behaves like an amorphous solid --- has large effective viscosity and is 
still characterized by the yield stress \cite{Yosh,voigt}. The 
solid-like state of lubricant corresponds to the shear stress $\sigma=0$, 
since in this case equation (\ref{eq1}) falls out of 
consideration\footnote{It will be shown further that non-zero 
stresses interval can meet the solid-like state of lubricant also.}. 
Equation (\ref{eq2}), containing the viscous stresses, is reduced to the 
Debye law representing the rapid relaxation of shear strain during 
microscopic time 
$\tau_{\varepsilon} \approx a/c \sim 10^{-12}$ s, where $a\sim 1$~nm is the 
lattice constant or the intermolecular distance and $c\sim 10^3$~m/s is the  
sound velocity. Thus, equation of heat conductivity (\ref{eq3}) assumes 
the form of simplest expression for temperature relaxation, which does 
not contain the terms representing dissipative heating and 
mechanic-and-caloric effect of viscous liquid. 

Equation~(\ref{eq2}) describes the flow of lubricant with velocity
$V=l\partial\varepsilon/\partial t$ due to action of appearing viscous shear
stress. Moreover, in accordance with Ref.~\cite{Aranson} in the absence of 
shear deformations the temperature mean-square displacement of 
molecules (atoms) is defined by equality $\langle u^2\rangle=T/Ga$. 
The average shear displacement is found from 
the relationship $\langle u^2\rangle =\sigma^2a^2/G^2$. The total mean-square 
displacement represents the sum of these expressions provided that the thermal 
fluctuations and the stress are independent. Above implies that the melting 
of lubricant is induced both by heating and influence of stress generated 
by solid surfaces in the course of friction \cite{Aranson}. 
It is shown \cite{Popov} that the plastic flow of 
lubricant layer is realized at presence of elastic stress. The action of shear 
stress results to reducing of shear modulus of lubricating material. 
Consequently, the friction force decreases with increasing 
velocity at the contact $V=l\partial\varepsilon/\partial t$ because the latter 
leads to the growth of the shear stress according to the Maxwell 
stress - strain $\varepsilon$ relationship: 
$\partial\sigma /\partial t {= -} 
\sigma / \tau_\sigma {+} G\partial\varepsilon/\partial t.$ 
It is assumed that the film becomes more liquid-like and the friction force 
decreases with the temperature growth due to decreasing activation 
energy barrier to molecular hops. 

\section{Hysteresis behavior}\label{sec:level3}

Let us consider the stationary states, at which the all derivatives in 
equations (\ref{eq1}) -- (\ref{eq3}) are equal to the zero, and values 
$\sigma, \varepsilon, T$ do not change in the lubricant. Then the first of 
these equations leads to the law similar to the Hooke one: 
\begin{equation}
\sigma = g(\varepsilon) \varepsilon, \quad g(\varepsilon) = 
g_\theta \left(1+\frac{\theta^{-1}-1}{1+(\varepsilon/\alpha)^\beta}\right),
\label{Hooke}
\end{equation}
where parameter $\theta = \Theta / G<1$, determining the ratio of the tilts 
for the deformation curve on the plastic and the Hookean regions, 
coefficients $g_\theta = \Theta / G_0 < 1$, and 
$\alpha = \varepsilon_p/\varepsilon_s$ are introduced. 

Dependence (\ref{Hooke}) at fixed $\alpha$, $g_\theta$, and $\theta$ is 
depictured in Fig. 1. It is apparent from here that two situations can be 
realized: at small $\beta$ the curve $\sigma(\varepsilon)$ increases 
monotonically (insert in the figure), and at 
\begin{eqnarray}
\beta > \frac{1 + \sqrt{\theta}}{1 - \sqrt{\theta}} 
\label{if}
\end{eqnarray}
it becomes non-monotonous. In the first case in accordance with the figure 
the continuous melting of lubricant takes place, in the second 
one --- lubricant melts abruptly at the 
increase of stresses to the point $A$, and here the 
transition to point $B$ occurs. At the further growth of the stresses the 
strain increases 
monotonically, and lubricant is liquid-like. 
If stress is decreased now to the point $C$ lubricant conserves liquid-like 
structure, and then it abruptly becomes solid during transition at point $D$. 
At the subsequent decrease of stresses lubricant is solid-like. 
Similar transitions are represented as the first-order phase transitions 
\cite{land_inst}, but between the states, which are not the pure  
thermodynamic phases. For understanding of these transformations the 
concept of shear melting is used \cite{Aranson}. Note that such 
hysteresis behavior was observed in the experiments \cite{exp1,Popov}. 

Using (\ref{Hooke}), it is possible to find the abscissas of transition 
points $A$ and $C$: 
\begin{eqnarray}
\varepsilon_{A,C} = 2^{-1/\beta}\alpha\left[ b(\beta - 1) - 2 \mp 
b\sqrt{(1-\beta)^2-4\beta / b}\right]^{1/\beta}, ~ b = \theta^{-1} - 1, 
\label{epsAC}
\end{eqnarray}
where the sign '-' corresponds to the point $A$, and sign '+' --- 
to the point $C$. 
From Eq. (\ref{epsAC}) it is seen that the length of jump at melting increases 
with growth $\alpha$, and the difference $\varepsilon_A - \varepsilon_C \to 0$ 
with the increase of $\beta$. Thus, at large $\beta$ 
(small $\alpha$) melting and solidification occur at 
the same value of strain ($\varepsilon_A \approx \varepsilon_C$) practically, 
but at 
different values of stresses $\sigma$. As well as in Refs. 
\cite{liq}--\cite{vestKIN}, we will accept as the order parameter the 
shear stress $\sigma$: at $\sigma > \sigma_A$ lubricant is liquid-like, and 
if $\sigma < \sigma_C$ it is solid-like. 
In the intermediate region $\sigma_C < \sigma < \sigma_A$ the state of 
lubricant is unstable, since it may exist in both phases. 

The dependence of the stationary shear stresses $\sigma_0, \sigma^m$ on 
the temperature of friction surfaces $T_e$ is presented in Fig.~2 at 
parameters of Fig. 1 (Fig. 2b meets the insert in Fig. 1). 
Apparently that it is non-monotonous, and in the interval 
$T_c^0<T_e<T_{c0}$ the two-valued section is realized inherent to 
first-order phase transitions. 
The dashed curve corresponds to the unstable stationary values of stresses 
$\sigma^m$, the solid curve -- to the stable $\sigma_0$. 
It is worth noting that $\sigma^m (T_e)$ meets the Hookean section of 
dependence $\sigma_0(\varepsilon_0)$. 

For further consideration it is necessary to write down expression for 
synergetic potential $V(\sigma)$. Within the framework of adiabatic 
approximation 
$\tau_\varepsilon, \tau_T \ll \tau_\sigma$ \cite{liq} it is possible to set 
$\tau_\varepsilon\dot\varepsilon\approx 0$, $\tau_T\dot T \approx 0$, and 
equations (\ref{eq2}), (\ref{eq3}) give 
\begin{eqnarray}
\varepsilon = \sigma - (2-T_e)\frac{\sigma}{1+\sigma^2},
\label{adiabat_eps} \\ 
T = T_e + (2-T_e)\frac{\sigma^2}{1+\sigma^2}.
\label{adiabat_T}
\end{eqnarray}
After substitution of expression (\ref{adiabat_eps}) in (\ref{eq1}) 
we obtain the Landau-Khalatnikov equation: 
\begin{equation}
\tau_\sigma\dot\sigma = 
-\frac{\partial V}{\partial \sigma}, \label{lanhal}
\end{equation}
where the synergetic potential is defined by equality 
\begin{equation}
V = \frac{\sigma^2}{2} - g_\theta\int\limits_0^\sigma\left[\sigma - 
(2-T_e)\frac{\sigma}{1+\sigma^2}\right]
\left[1+\frac{\theta^{-1}-1}{1+\left(
\sigma/\alpha - \sigma(2-T_e)/(\alpha+\alpha\sigma^2)
\right)^\beta}\right] d\sigma.
\label{potential}
\end{equation}
Note that in the region $T_{cA}<T_e<T_{cC}$ this potential does not give 
correct result, since here the model has the unstable solution, describing the 
decrease of stresses with growth of strain, and this does not take into 
account hysteresis. For finding the form of potential in the indicated region 
we replace (\ref{eq1}) by equation: 
\begin{equation}
\tau _{\sigma}\dot{\sigma}=-\sigma + \delta, \label{eq1_new}
\end{equation}
where $\delta$ is the value of the conserved stresses. In the corresponding 
Landau-Khalatnikov equation (\ref{lanhal}) the potential is 
fixed by expression: 
\begin{equation}
V^\prime = \frac{\sigma^2}{2} - \delta\sigma. 
\label{potential_stat}
\end{equation}
It is seen that $V^\prime$ depends only on $\sigma$, i.e. in the situation, 
where stress is conserved at the change of temperature, the value 
$V^\prime$ remains constant. 

The dependence of the potential (\ref{potential}) on the stress value 
at the fixed temperatures of friction surfaces is presented in Fig.~3 
corresponding to the parameters of Fig. 2a. 
The hysteresis, shown in Fig. 1, is realized in this situation. 
We will consider this case in more detail. 

Below than the critical value $T_c^0$ stresses are absent in lubricant 
($\sigma_0=0$). The form of potential, shown by curve 1 in Fig. 3, 
corresponds to this interval of temperatures. 
Here one zero minimum is realized at $\sigma_0=0$, thus lubricant 
is solid-like. 
At point $T_e = T_c^0$ a plateau appears on the dependence 
$V(\sigma)$ (curve 2). 
With the further increase of temperature in a region $T_c^0 < T_e < T_{c0}$ 
the potential has the form shown by curve 3. Here the potential 
barrier appears, which separates the zero and non-zero minimums of potential. 
In connection with it the system can not come in the stable state 
$\sigma_0 \ne 0$, and the zero value of shear stress $\sigma$ is 
realized. The dashed curve in Fig. 2a corresponds to the maximum of potential, 
the solid curve --- to its non-zero minimum. 
As is apparent from figure at $T_e=T_{c0}$ the jump-like increase 
of value $\sigma$ takes place, and the system turns to the section of 
dependence 
$\sigma_0(T_e)$ (point $A^\prime$). This transition is caused by that 
at $T_e=T_{c0}$ the maximum $V(\sigma)$ disappears, and with subsequent 
growth of $T_e$ the one non-zero minimum of potential is realized 
(curve 4 in Fig. 3). The 
expression for the critical temperature $T_{c0}$ is obtained from the 
condition $\partial V/ \partial \sigma = 0$, 
where $V$ is the synergetic potential (\ref{potential}): 
\begin{equation}
T_{c0} = 1+ \theta /g_\theta \equiv 1 + G_0/G. \label{Tc0}
\end{equation}
However, section $AC$ is unstable, because here the stress is decreased at 
growth of strain ($A$, $B$, $C$ and $D$ meet the similar points 
in Fig. 1). In this connection further the system goes on the way 
$A^\prime$ -- $B^\prime$ (this transition is already described by 
potential $V^\prime$ (\ref{potential_stat}), since stress is conserved), 
and passes to the plastic flow section, which corresponds to the 
liquid-like structure of lubricant. With the subsequent increase of 
temperature $T_e$ the 
stationary value of stresses $\sigma_0$ grows, and lubricant becomes 
less viscous, here $V(\sigma)$ has the form shown by curve 4. 
If now the temperature of friction surfaces is decreased, the 
lubricant is liquid-like to the value $T_{cC}$, further the stress is 
conserved to the critical $T_e = T_{cA}$ (corresponding to the point 
$A$)\footnote{In general case the system has to move along the curve $CA$, 
but it is unstable. Therefore it is necessary to introduce the 
described hysteresis, and to take into account that at this transition 
the system is described by potential $V^\prime(\sigma)$ 
(\ref{potential_stat}).}. As is seen from the insert in the picture at 
$T_e = T_{cA}$ the system passes to the point 
$A$, because here the dependence $\sigma_0(T_e)$ becomes stable. At this 
transition system of somehow ''jump over'' the potential barrier 
(dashed curve of dependence). 
It is related to that before the jump the system is described by other 
potential $V^\prime$, which is characterized by absence of barrier. Here the 
solidification of lubricant occurs, because transition takes place on the 
stable part 
of Hooke section (according to Fig. 1 temperature $T_e<T_{cA}$ corresponds to 
the Hooke domain). Now, the system is described by potential $V(\sigma)$ 
(\ref{potential}) with the barrier shown by curve 3 in Fig. 3. 
With the further decrease of the temperature of friction surfaces at the 
point $T_e=T_c^0$ stresses decrease abruptly to the zero, since the 
barrier disappears 
and there is one zero minimum of $V(\sigma)$. This situation corresponds 
to the solid-like state of lubricant also, but with the zero value of stress. 
It is supposed that the solid-like states at $\sigma_0=0$ and 
$\sigma_0 \ne 0$ differ by structure, because of transition between them 
takes place according to the mechanism of the first-order phase transition. 
Thus, the solid-like structure of lubricant at a temperature below $T_c^0$ 
is similar to the solid state, and the solid-like structure above 
indicated temperature has the signs of the liquid state, but lubricant 
behaves as a solid-like on the whole. 
The different solid-like states correspond to the different modes of friction 
in the studied system \cite{Braun}. 
At further transition to the liquid mode of friction the viscosity of 
lubricant (\ref{visc}) decreases, and it flows. 

At described transitions the stationary values of stresses are conserved in 
intervals $T_{cA^\prime}<T_e<T_{cB^\prime}$ and $T_{cA}<T_e<T_{cC}$. Obviously, 
that equality $\sigma_0 = {\rm const}$ is satisfied with the increase of 
temperature at $T_{cA^\prime}<T_e<T_{cB^\prime}$, since it is necessary to 
give some energy to lubricant for melting. Conservation of stresses 
in the region $T_{cA}<T_e<T_{cC}$ at decrease of the sheared surfaces 
temperature takes place, because for transition of lubricant in the solid-like 
state it has to return energy. It is worth noting that at the conservation of 
stresses between solid-like and liquid-like phases of lubricant, it is in the 
intermediate state with differing from them structure.

The situation, which is given by the insert to the figure 1, is 
shown in Fig. 2b when hysteresis is not realized in coordinates 
$\sigma_0(\varepsilon_0)$. This corresponds to the more simple case described 
in work \cite{liq}. Below the value $T_{c0}$ we have the solid-like state 
of lubricant ($\sigma_0=0$), at $T_e = T_{c0}$ it melts at transition to 
the plastic flow region, and with the subsequent increase of $T_e$ it 
becomes more liquid. Then, with the decrease of temperature to the value 
$T_c^0$ lubricant is liquid-like ($\sigma_0\ne 0$), at $T_e = T_c^0$ it 
solidifies abruptly ($\sigma_0=0$). 

The distinctive feature of such behavior is that whole Hooke domain 
is unstable, since it meets the maximum of potential. There is only 
one type of solid-like state of lubricant with the zero value of stress. 

Depending on the parameters of the system there the second-order phase 
transition can be observed, when the temperature $T_{c0}$ is moved on the 
left-hand side from $T_c^0$. Indeed, the unstable stationary values of 
stresses $\sigma^m$ are not realized, 
and hysteresis disappears, which is characterized by a presence of $T_c^0$. 
In this case the potential, shown by curve 1 in Fig. 3, corresponds to the 
solid-like structure of lubricant, which is transformed to the form 
represented by curve 4 with the increase of temperature. Thus, the continuous 
transformation takes place, since the potential barrier is absent. 
Here it is also possible to select two situations -- when hysteresis of 
$\sigma_0(\varepsilon_0)$ dependence is observed, 
and without it \cite{thesys}. 

\section{Phase kinetics}\label{sec:level4}

In accordance  with the experimental data for organic lubricating materials 
\cite{Yosh,pla} the relaxation time of stresses at normal pressure is 
$\tau_{\sigma}\sim 10^{-10}$~s. Since ultrathin lubricant film has less 
than four molecular layers, temperature relaxes to the value $T_e$ during 
time satisfying the condition: 
\begin{equation}
\tau_T \ll\tau_\sigma,~\tau_\varepsilon. 
\label{time_rel}
\end{equation}
According to this we will assume in equation (\ref{eq3}) 
$\tau_T\dot T \approx 0$. Also for convenience the time is measured 
in units of $\tau_\sigma$. 
As a result, we will obtain the two-parameter system in the form: 
\begin{eqnarray}
&&\dot{\sigma}=-\sigma + g(\varepsilon)\varepsilon, \label{eq1_new1} \\
&&\tau\dot{\varepsilon}=-\varepsilon + 
(T_e-1-\sigma\varepsilon+\sigma^2)\sigma, \label{eq2_new1}
\end{eqnarray} 
where $\tau \equiv \tau_\varepsilon/\tau_\sigma$. 
The corresponding phase portraits are presented in Figs.~4 -- 6 at the 
parameters of figure 2a for different $T_e$. 

The phase portraits describing the behavior of lubricant in the solid-like 
state ($T_e < T_c^0$) are shown in Fig.~4 for different ratios of relaxation 
times. 

Particularly, Fig.~4a corresponds to the value $\tau=0.01$. 
The isoclines obtained at equating to zero of derivatives in the 
equations (\ref{eq1_new1}) and (\ref{eq2_new1}) are shown by dashed curves 
1 and 2, respectively. 
Thus, curve 1 meets the system parameters at which stresses do not change, 
and line 2 corresponds to the case of strain conservation. These lines 
intersect at the origin of coordinates forming the unique stationary point 
$D$, which is a node. It is apparent that phase trajectories converge to the 
node $D$, i.e. stresses relax to the zero value. 
Thus, at motion on a phase plane at the arbitrary initial conditions 
there are two stages: on the first one, the instantaneous relaxation of the 
system takes place to the line near to isocline 2, on the second one 
--- the slow motion on the indicated curve. At the first stage the stresses 
are conserved that reminds the transition between the friction 
modes described in section~\ref{sec:level2}. 
Apparently, on the final stage of system relaxation to the stationary 
point $D$ the values of stress and strain go out in the negative domain. 
This can be interpreted as reversible motion at which the top wall of friction 
moves in reverse direction. 
Thus, stresses change the sign in opposite one, because of the direction of 
motion changes, and strain becomes negative also. 
On the other hand, it is possible to neglect the negative region, 
considering it as has no physical meaning. We can assume that at achievement 
by strain the zero value the system abruptly goes to the origin of 
coordinates, and the equilibrium sets in \cite{vestKIN}. 

The phase portrait shown in Fig.~4b is built for the case when relaxation 
times of stresses and strain coincide $(\tau=1)$. It is also characterized by 
the singular point $D$, representing the node around which the weakly 
pronounced oscillations of short durations are realized now till setting of 
equilibrium. 
Here the cases are possible when stress $\sigma$ at first increases, and 
then decreases, and reverse situation. It means that to that moment, 
when the system will come to the equilibrium (the origin of the 
coordinates), the stick-slip motion is possible. 
For example, in accordance with phase trajectories, which begin at $\sigma=0$, 
the lubricant is solid-like at first (stresses are equal to zero), then it 
begins to melt (stresses increase), and then it solidifies again (at setting 
of equilibrium). In order to avoid misunderstanding let us turn attention to 
the following circumstance. According to phase trajectories the cases are 
possible when with the growth of stresses the strain decreases. 
Above for the stationary 
values of stresses it was interpreted as instability resulting in hysteresis. 
Now, the rapid motion is realized. It allows us 
to assume that system is in the unstable state, and 
oscillation mode is realized in which the decrease of $\varepsilon$ is 
possible with the increase of $\sigma$. 

Figure~4c meets the case $\tau=100$. Here, as well as in Fig.~4a, the two 
stages are selected: the rapid relaxation to the line near to isocline 1, 
and further the slow motion along it. At the first stage the strain changes 
weakly, and stresses decrease very rapidly, if their initial values are on 
the right-hand side from isocline 1, or they increase at initial $\sigma$ on 
the left-hand side from it. At the second stage in the top 
part of phase portrait the configuration point moves along the plastic 
section, and in bottom one -- along the Hooke domain. 
During the way of this point between the maximum and 
minimum of isocline 1 stresses increase with decrease of strain. The motion 
is slow here, and therefore the hysteresis has to be observed similar to 
described in Fig. 1. This feature is not shown in phase portrait, and its 
study requires the additional analysis, because there a lot of limiting cases 
can arise, which are not considered within the framework of the offered work. 

The phase portraits in Fig.~5 are presented for the same parameters and ratios 
of relaxation times as in Fig.~4. But they are built for the temperature 
corresponding to the section of Fig. 2a, for which stable and unstable 
stationary values of shear stresses are realized ($T_c^0<T_e<T_{c0}$). 
In this case the system potential has the form shown by curve 3 in Fig. 3. 
As well as above, lines 1 and 2 are isoclines. 
The phase portraits are characterized by three singular points: 
by the node $D$ at the origin of coordinates, which describes the dry friction; 
by the saddle $N$ corresponding to the maximum of dependence $V(\sigma)$ 
(the unstable stationary point); 
by the node $O$ meeting the non-zero stationary stress, which corresponds 
to the unstable section $AC$ of dependence $\sigma_0(T_e)$ shown in Fig. 2a. 
These points are given by intersections of isoclines. Depending on the initial 
conditions the system may come as a result of relaxation both to the mode of 
stable dry friction (node $D$) and to the above described unstable section 
(node $O$). 

At $\tau = 0.01$ there is the picture shown in Fig.~5a. Here, as well as in 
Fig.~4a, phase trajectories rapidly converge to the line near to isocline 2 
from any point of phase plane at conservation of stresses. Further the system 
relaxes to the nodes $D$, or $O$, and determined by these points the 
stationary modes of friction set in. The inclination of curve, along which 
motion occurs during the second stage, depends on initial conditions. 
So, the system relaxes to the point $O$ along the plastic section of 
isocline 2, to the point $D$ -- along its Hookean one. 
Note that in the course of time lubricant becomes more liquid, if 
$\sigma_N<\sigma_{i}<\sigma_O$, and vice versa more viscous at 
$\sigma_{i}>\sigma_O$, where $\sigma_{i}$, $\sigma_N$, and 
$\sigma_O$ are the initial and the stationary values of stresses. 
In these cases the system comes to the singular point $O$. 
At $\sigma_{i}<\sigma_N$ the lubricant solidifies during time, and 
dry friction is realized (point $D$). 

As is obvious from Fig.~5b, for $\tau = 1$ the different types of stick-slip 
motion are possible at setting of equilibrium values of stress and strain. 

The phase portrait is shown in Fig.~5c for $\tau=100$, where, as well as 
above, there are two stages. Since the stationary point $O$ is on the unstable 
section, for the parameters of this figure the hysteresis is characteristic. 

The phase portraits for the temperature domain of figure 2a, corresponding 
to the stable sliding friction ($T_e>T_C$), are shown in Fig.~6. Here in the 
course of evolution the non-zero stationary value of shear stresses is 
set $\sigma_0\ne 0$ meeting the minimum of synergetic potential 
$V(\sigma)$ (the maximum of distribution function of stresses $P(\sigma)$ 
over their value). The phase portraits are characterized by two singular 
points --- by the saddle $D$ at the origin of coordinates and the node $O$ at 
the non-zero values of stress and strain, which are given by intersections 
of isoclines 1 and 2. 
 
At $\tau = 0.01$ the situation is observed shown in Fig.~6a. Here, as well as 
above, phase trajectories converge rapidly to the line near to isocline 2 
from any point of phase plane at conservation of stresses. Then, the system 
relaxes to the non-zero value $\sigma_0\ne 0$, as a result, the stationary 
sliding friction sets in. However, the curve, along which motion occurs 
during the second stage, corresponds to the plastic section of 
dependence $\sigma(\varepsilon)$, i.e. system is liquid-like always, 
except only the cases when initial value of 
stresses is near zero (the melting occurs). Note that in the course of 
time the lubricant becomes more liquid, if $\sigma_{i}<\sigma_O$, and vice 
versa --- more viscous at $\sigma_{i}>\sigma_O$. 

As is apparent from Fig.~6b, for $\tau = 1$ at setting of stationary sliding 
friction the different types of stick-slip motion are possible. However, at 
any initial conditions the system comes to the steady sliding friction. 

The phase portrait is shown in Fig.~6c for $\tau=100$, where, as well as in 
Figs.~4c and 5c the two above described stages are seen. The stationary point 
$O$ is on the plastic flow region. The hysteresis is characteristic 
also for the parameters of this figure. 

The difference of described in Figs. 4, 5, and 6 situations is achieved due 
to the change of isocline 2 form with the variation of temperature $T_e$ of 
friction surfaces, while the form of isocline 1 does not depend on it. 
The value of $\beta$ influences substantially on the phase portraits, 
however, its variations do not change qualitatively the character of the 
system behavior. 
At $\tau \gg 1$ it is necessary to introduce hysteresis \cite{thesys,exp2}, 
because the slow motion of the system occurs near the isocline 1, and the 
condition of stationarity is valid in every time moment. The hysteresis is 
similar to described in Fig. 1, since the dependence, presented in this 
figure, represents isocline 1 shown in phase portraits. 

\section{Conclusion}\label{sec:level6}

The above consideration shows that the hysteresis, realized at melting of 
thin lubricant film according to the mechanism of first-order phase 
transition, can be described taking into account the deformational defect of 
the shear modulus. The basic feature of such behavior is that lubricant melts 
and solidifies at the different values of shear stress, which acts as the 
order parameter. The two solid-like states of lubricant and one liquid-like 
phase have been found, the transition between which takes place in accordance 
with indicated hysteresis. 
The phase kinetics is studied, and depending on initial conditions the 
different types of stick-slip motion are predicted. The elastic and 
thermal parameters of lubricant are defined, at which as a result of this 
motion the sliding, or the dry friction sets in. 

\section*{Acknowledgements}

We express our gratitude to the organizers and participants of Working 
conference-seminar of Institute for Condensed Matter Physics 
of National Academy Sciences of Ukraine (2-3 of June, 2005, Lvov) within 
the framework V Ukrainian competition of young scientists in the field of 
statistical physics and theory of condensed medium for discussing and 
supporting of the work. We thank also Dr. A. Kiselev and Dr. S. Lukjanets 
for the discussion of work. The work was partly supported by a grant of 
the cabinet of Ukraine.

\newpage
\begin{center}
{\bf Figure captions}  \
to the work of A.V.~Khomenko, I.A.~Lyashenko "Hysteresis phenomena at 
ultrathin lubricant film melting in the case of first-order phase transition" 
\end{center}

\begin{description}

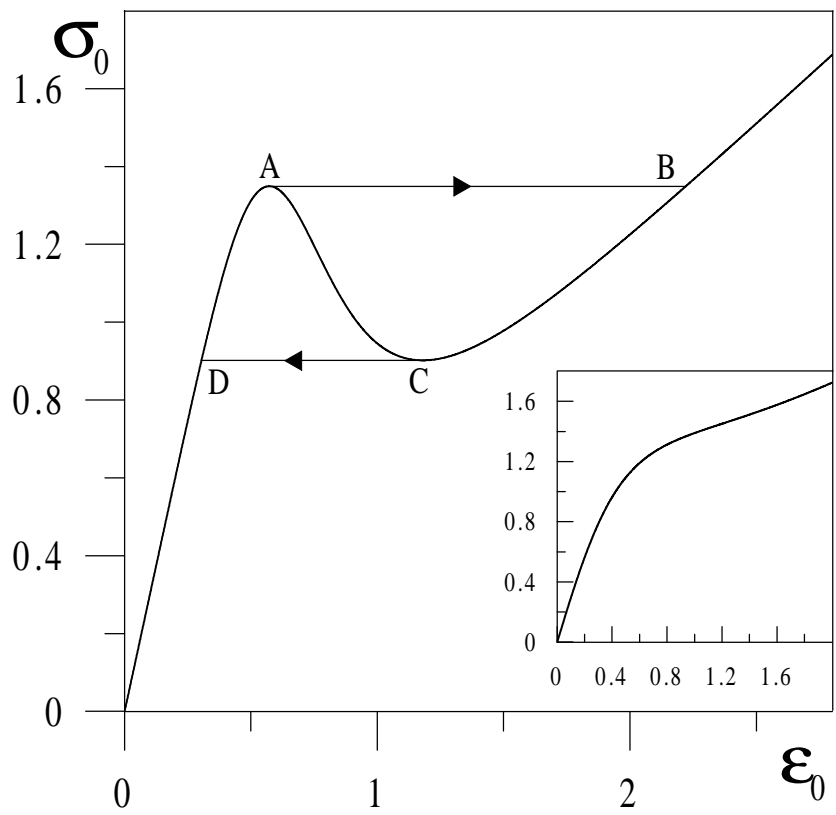
\item[Fig.~1.] The dependence of the stationary values of the shear 
stresses $\sigma_0$ on the strain $\varepsilon_0$ (\ref{Hooke}) at 
$\theta = 0.2$, $g_\theta = 0.6$, 
$\alpha = 0.7$, $\beta = 5.0$ (in the insert $\beta = 2.0$). 

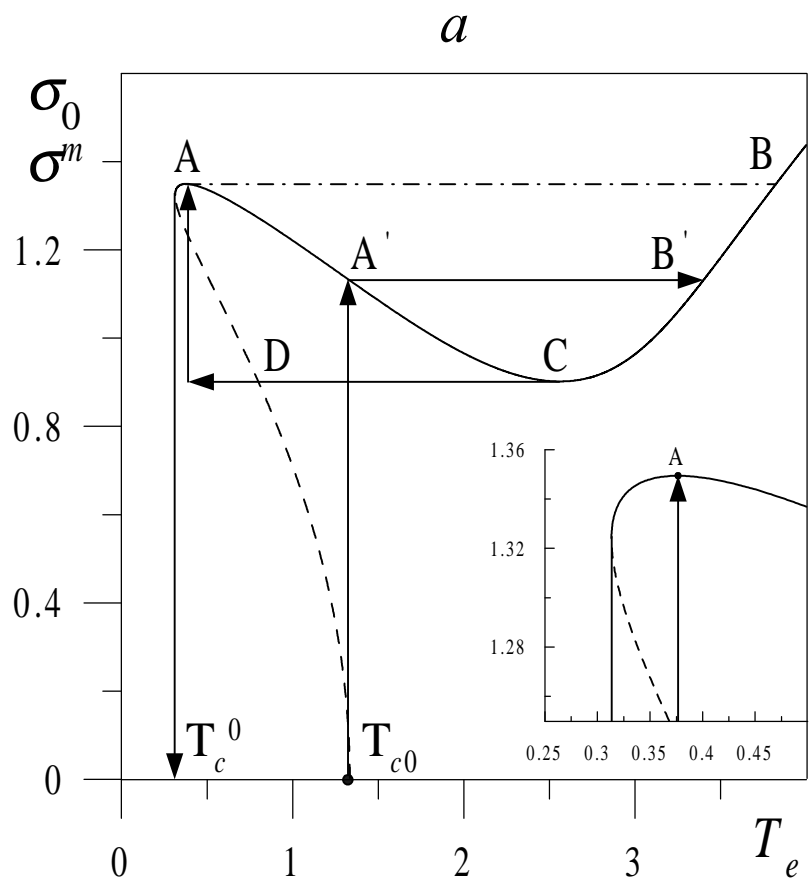
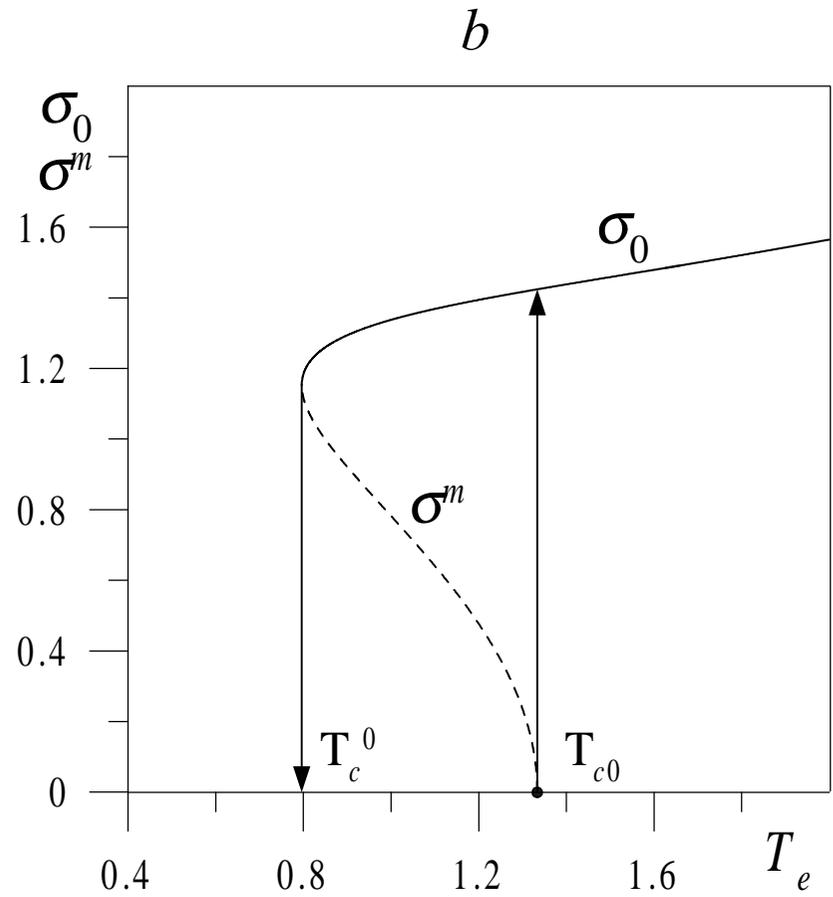
\item[Fig.~2.] The dependence of the stationary values of the shear stresses 
$\sigma_0, \sigma^m$ on the temperature of friction surfaces $T_e$ at 
parameters of Fig.~1 (Fig. 2b corresponds to the parameters of insert 
in Fig. 1). 

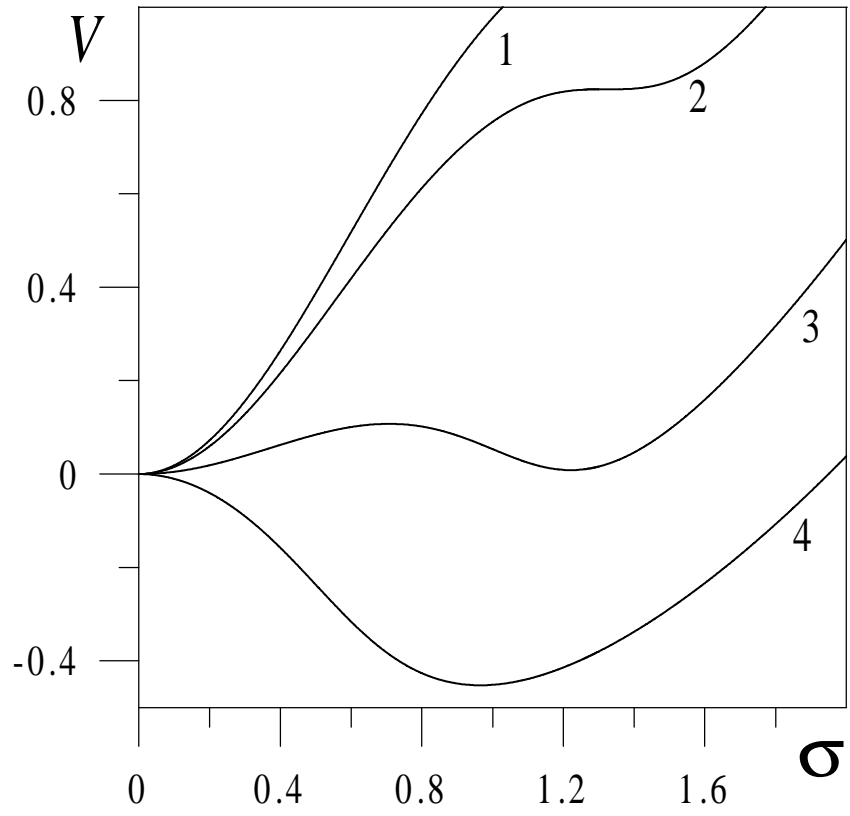
\item[Fig.~3.] The dependence of the synergetic potential $V$ 
(\ref{potential}) on the shear stresses $\sigma$ at parameters of 
Fig.~2a, curves 1--4 correspond to the values of temperature 
$T_e = 0.1, 0.3125, 1.0, 2.0$, respectively. 
            
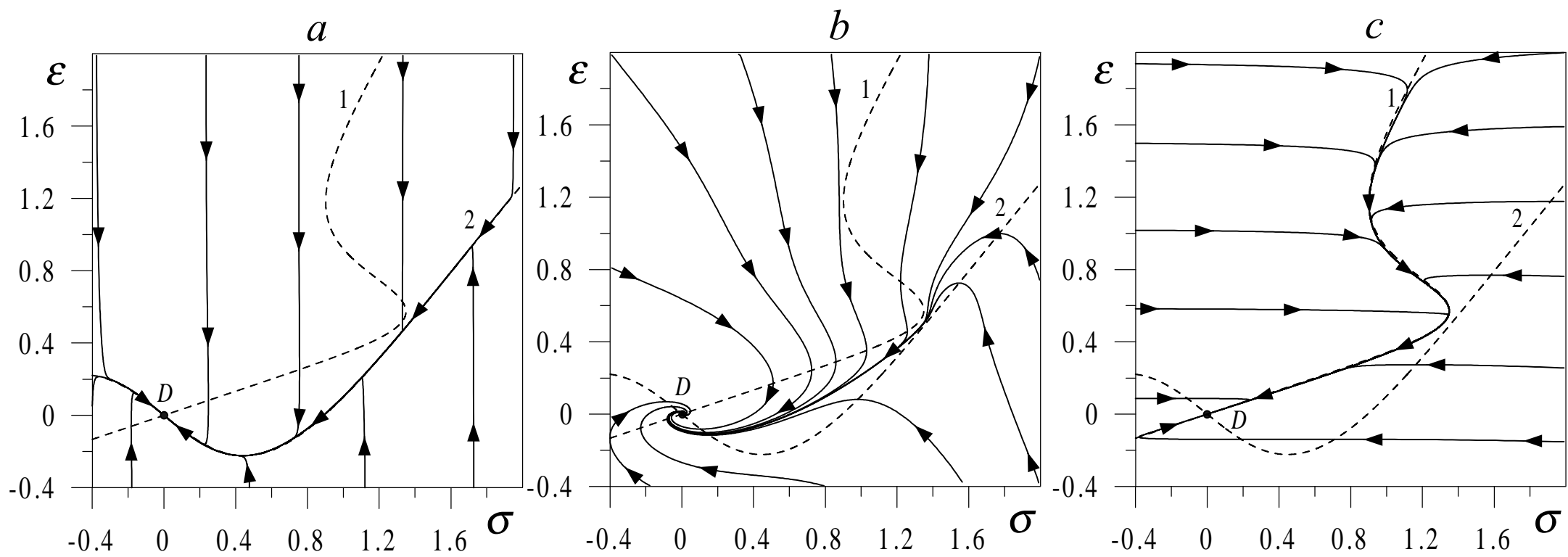
\item[Fig.~4.] The phase portraits at parameters of Fig.~2a and $T_e=0.2$: 
(a) $\tau_T\ll\tau_\varepsilon = 0.01\tau_\sigma$; 
(b) $\tau_T\ll\tau_\varepsilon = \tau_\sigma$; 
(c) $\tau_T\ll\tau_\varepsilon = 100\tau_\sigma$. 

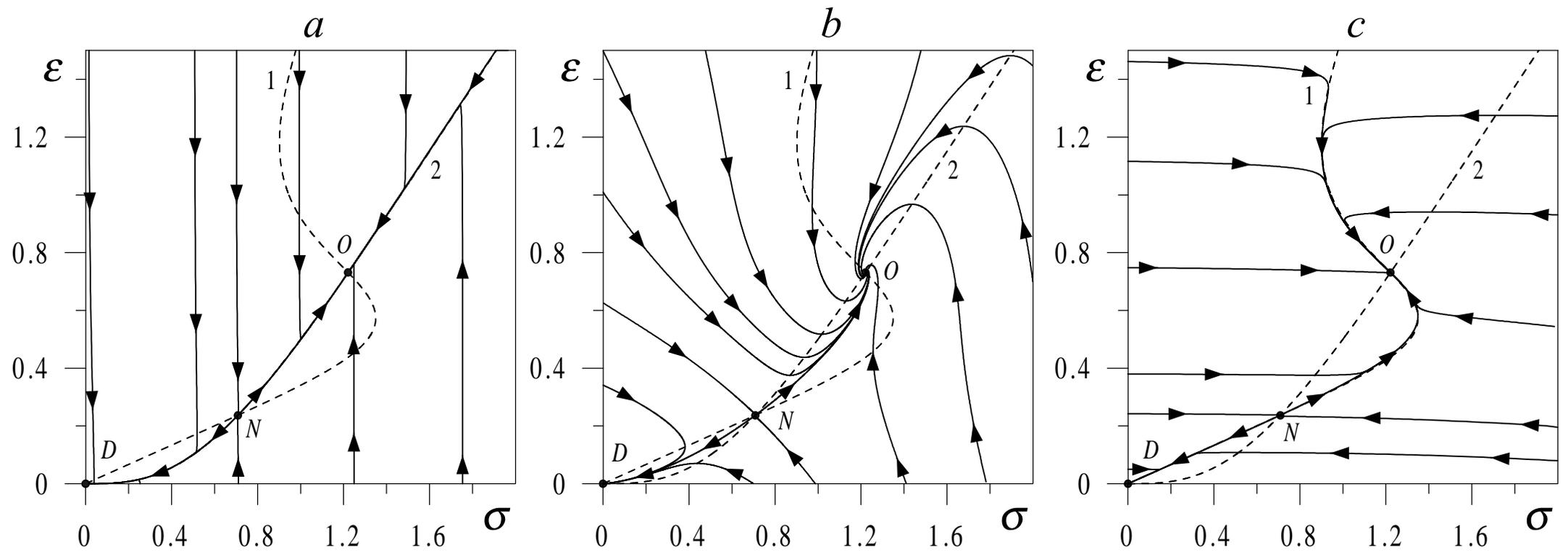
\item[Fig.~5.] The phase portraits at parameters of Fig.~2a and $T_e=1.0$: 
(a) $\tau_T\ll\tau_\varepsilon = 0.01\tau_\sigma$; 
(b) $\tau_T\ll\tau_\varepsilon = \tau_\sigma$; 
(c) $\tau_T\ll\tau_\varepsilon = 100\tau_\sigma$. 

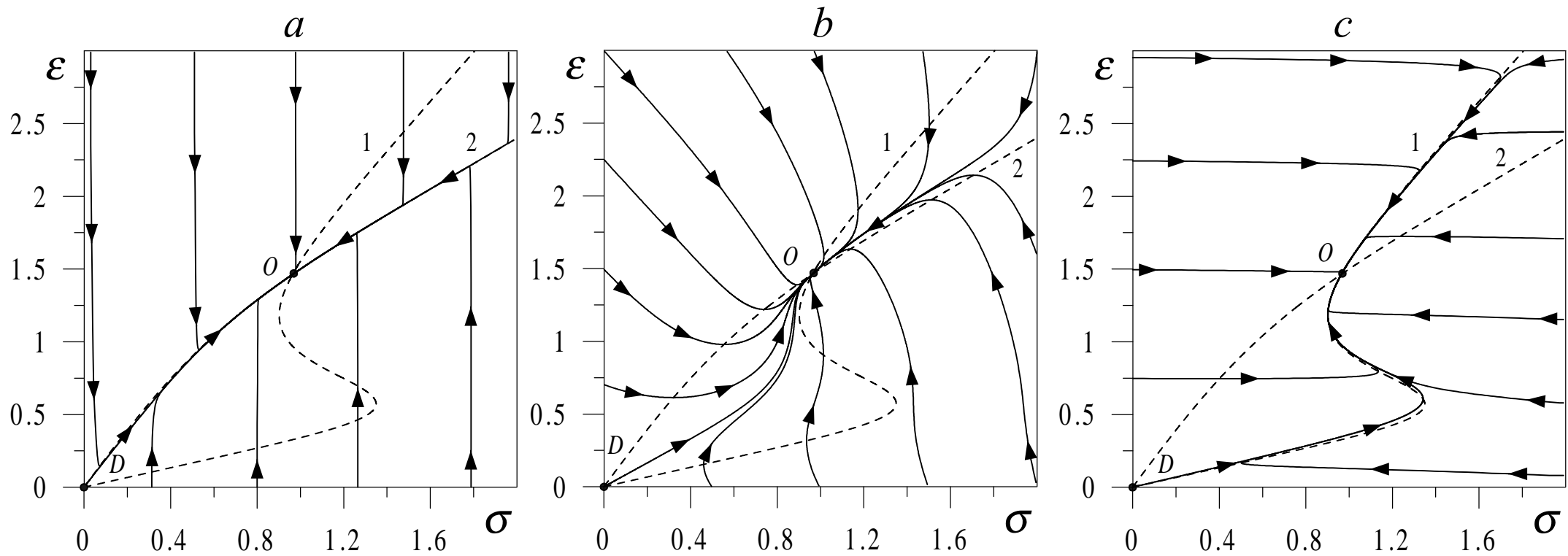
\item[Fig.~6.] The phase portraits at parameters of Fig.~2a and $T_e=3.0$: 
(a) $\tau_T\ll\tau_\varepsilon = 0.01\tau_\sigma$; 
(b) $\tau_T\ll\tau_\varepsilon = \tau_\sigma$; 
(c) $\tau_T\ll\tau_\varepsilon = 100\tau_\sigma$. 

\end{description}

\end{document}